\input harvmac

\def\eg{{\it e.g.,}\ }
\def\ie{{\it i.e.,}\ }
\def\D{{\cal D}}

\def\({\left (}
\def\){\right )}
\def\[{\left [}
\def\]{\right ]}
\def\ssc{\scriptscriptstyle}
\def\A{A_{\ssc H}}
\def\buildchar#1#2#3{{\null\!
   \mathop{\vphantom{#1}\smash#1}\limits%
   ^{#2}_{#3}%
   \!\null}}

%
\Title{\vbox{\baselineskip12pt
\rightline{gr-qc/9705065 \hfill McGill/97-09}}}
{\vbox{\centerline{Pure states don't wear black}}}

\baselineskip=12pt
\centerline {Robert Myers\footnote{$^1$}{Internet: 
rcm@hep.physics.mcgill.ca}}
\medskip
\centerline{\sl Department of Physics}
\centerline{\sl McGill University}
\centerline{\sl Montr\'eal, Qu\'ebec H3A-2T8 Canada}

\bigskip
\centerline{\bf Abstract}
\medskip
\baselineskip = 20pt 
Recently, string theory has provided some remarkable new
insights into the microphysics of black holes. I argue
that a simple and important lesson is also provided with
regards to the information loss paradox, namely, pure quantum states
do not form black holes! Thus it seems black hole formation, as
well as evaporation, must be understood within the framework of
quantum decoherence.

\Date{}

\def \pr  {Phys. Rev. }

\def \cmp {Commun. Math. Phys.}

\gdef \jnl#1, #2, #3, 1#4#5#6{ {\it #1~}{\bf #2} (1#4#5#6) #3}

\lref\rad{S.W.~Hawking, \jnl \cmp, 43, 199, 1975.}
\lref\radb{S.W.~Hawking, \jnl \pr, D14, 2460, 1976.}
\lref\conflict{see for example: J.~Preskill, 
``Do black holes destroy information,'' preprint hep--th/9209058.}
\lref\beken{J.D.~Bekenstein, \jnl \pr, D7, 2333, 1973;
{\bf D9} (1974) 3292.}
\lref\dual{J. Polchinski, ``String Duality: A Colloquium,''
e-print hep-th/9607050;
C. Vafa, ``Lectures on Strings and Dualities,''
e-print hep-th/9702201.}
\lref\dbran{J. Polchinski, ``TASI Lectures on D-branes,''
e-print hep-th/9611050;
J. Polchinski, S. Chaudhuri and C.V. Johnson, ``Notes on D-branes,''
e-print hep-th/9602052.}
\lref\first{A.~Strominger and C.~Vafa,
{\it Physics Letters} {\bf B379} (1996) 99
[hep-th/9601029].}
\lref\four{C.V. Johnson, R.R. Khuri and R.C. Myers,
{\it Phys. Lett.} {\bf B378} (1996) 78 [hep-th/9603061];
J.M. Maldacena and A. Strominger, 
{\it Phys. Rev. Lett.} {\bf 77} (1996) 428 [hep-th/9603060].}
\lref\near{C.G.~Callan and J.M.~Maldacena, 
{\it Nucl. Phys.} {\bf B472} (1996)
591 [hep-th/9602043];
G.T.~Horowitz and A.~Strominger, 
{\it Phys. Rev. Lett.} {\bf 77} (1996) 2368 [hep-th/9602051].}
\lref\donc{G.T. Horowitz and D. Marolf, {\it Phys. Rev.} {\bf D55}
(1997) 3654 [hep-th/9610171].}
\lref\scat{S.R.~Das and S.D.~Mathur, {\it Nucl. Phys.}
{\bf B478} (1996) 561 [hep-th/9606185];
{\it Nucl. Phys.} {\bf B482} (1996) 153 [hep-th/9607149];
J.~Maldacena and A.~Strominger, {\it Phys. Rev.} {\bf D55} (1997)
861 [hep-th/9609026].}
\lref\garya{G.T. Horowitz and D. Marolf, {\it Phys. Rev.}
{\bf D55} (1997) 846 [hep-th/9606113];
{\it Phys. Rev.} {\bf D55} (1997) 835 [hep-th/9605224].}
\lref\garyb{G.T.~Horowitz and H.-S.~Yang, ``Black Strings and Classical
Hair,'' e-Print hep-th/9701077.}
\lref\law{J.M.~Bardeen, B.~Carter and S.W.~Hawking, \jnl \cmp,
31, 161, 1973.}
\lref\uni{S.R.~Das, G.W.~Gibbons and S.D.~Mathur, {\it Phys. Rev. Lett.}
{\bf 78} (1997) 417 [hep-th/9609052].}
\lref\bright{
N.~Kaloper, R.C.~Myers and H.~Roussel, ``Wavy Strings: Black or Bright?'',
to appear in {\it Physical Review D}, e-Print hep-th/9612248;
A.A.~Tseytlin, {\it Mod. Phys. Lett.} {\bf A11} (1996) 689
[hep-th/9601177].}
\lref\tseytb{A.A.~Tseytlin, {\it Phys. Lett.} {\bf B363} (1995) 223
[hep-th/9509050].}
\lref\dec{W.H. Zurek, {\it Physics Today} {\bf 44}, No. 10 (1991) 36.}

Over twenty years ago, Stephen Hawking\rad\ made the remarkable
discovery that quantum mechanics leads to the emission of thermal
radiation from classical black holes. 
A vexing puzzle concerning the compatibility of quantum theory
and general relativity was raised shortly afterwards by
the question: What is the final end state which is produced 
when a pure quantum state collapses to form a black hole which
then emits thermal radiation?
Hawking\radb\ proposed that the black hole continues radiating
until, having radiated away its entire mass, it disappears,
leaving behind only thermal radiation 
in a mixed quantum state. In this progression from a pure to
a mixed state, the details of the original state are lost and
unitary time evolution, a basic tenet
of quantum theory, is violated. Hence this proposal suggests that the
basic principles of quantum physics, while successful
in describing everyday experiments, cannot accomodate gravity
without radical revision.
Attempts to produce a less distressing scenario in which information
of the original state is preserved 
have proven unsatisfactory\conflict,
and so this disturbing puzzle, the information paradox, remains unsolved.

A related puzzle arises because combining Hawking's result with the classical
laws of black hole mechanics\law\ indicates that, as anticipated by
Bekenstein\beken, black holes have an intrinsic entropy
proportional to the surface area of their
horizons\ref\area{This result also applies
for Einstein gravity in higher spacetime dimensions,
in which case the surface area $\A$ refers to the volume of a space-like
cross-section of the horizon, \eg in five dimensions,
$\A$ becomes a three-dimensional volume.}:
\eqn\entropy{S_{\ssc BH}={A_{\ssc H}/4G}\ \ .}
In this context, one has a thermodynamic understanding of this
entropy, \ie it is related to energy unavailable for work. 
A longstanding problem has been finding a statistical 
mechanical understanding of this entropy in terms of some
microscopic degrees of freedom. Recently, superstring theory
has provided some extraordinary new insights into this question.

This progress ensued from the realization that in string theory
extended objects other than strings also play an important role.
In particular, Dirichlet branes, or
D-branes\dbran, form a diverse class of extended
objects, distinguished by a simple description in the framework
of perturbative or weakly-interacting strings.
Yet D-branes exhibit a rich dynamics, including
a wide variety of complicated bound states. Characterizing
these bound states by certain asymptotic features,
\eg mass and charges, the degeneracy $\D$ of these configurations 
is amenable to statistical mechanical analysis
within the perturbative regime. If the string coupling constant 
(and hence also Newton's constant for the theory)
is increased, it is believed that the bound states undergo
gravitational collapse to form black holes.
In this complementary strong-coupling regime, one can use the
low-energy string equations --- essentially Einstein gravity
coupled to various massless fields --- to determine the corresponding
black hole solution, for which one can calculate
the Bekenstein-Hawking entropy \entropy.
Given appropriate arguments ({\it e.g.,} supersymmetry) that the degeneracy
is invariant when the coupling changes, one might expect
$S_{\ssc BH}=\ln \D$. These calculations were first
elucidated for a class of extremally charged black holes
in five dimensions\first, and yielded a striking agreement between
$S_{\ssc BH}$ and the statistical entropy.
This analysis was quickly extended to a variety of other configurations,
including black holes in four dimensions\four\ and slightly
non-extremal black holes\near. Again a precise
agreement of the entropies was found in each case.

Suprisingly, it was further found that the D-branes described the
dynamics of near-extremal black holes, at least at low energies\scat.
In particular, it was shown that the black hole decay
rates and absorption cross-sections were precisely reproduced.
As a simple
example, the absorption cross-section for a scalar (with energy $\omega$)
incident on an extremal D-brane configuration is
\eqn\cross{
\sigma_{\rm abs}=A_{\ssc H}\,\beta\omega\,(\langle n_{\omega/2}\rangle+1) }
where $A_{\ssc H}$ is the quantity that matches to the horizon area,
while $\beta$ is a scale 
which characterises the density of states of the D-brane
configurations. Finally, $n_{\omega/2}$
is the number of internal excitations with energy $\omega/2$ in a given
configuration. To good accuracy,
$\langle n_{\omega/2}\rangle$ averaged over all of the
degenerate D-brane configurations is approximated by an effective
canonical ensemble
\eqn\can{\langle n_{\omega/2}\rangle={1\over e^{\beta\omega}-1}\ .}
This result precisely reproduces the cross-section for the
corresponding extremal black hole\scat.
In particular for small frequencies ($\beta\omega<<1$), one
has the finite result $\sigma_{\rm abs}=A_{\ssc H}$, a
universal result for black holes\uni.
Thus the D-branes give a very robust model
of near-extremal black hole dynamics.

Given that the D-branes provide such a successful model of black
hole microphysics, one might ask if there are any lessons to be learned
with regard to the information paradox. Here I 
argue that they provide a simple but important message:
pure quantum states do not form black holes!

First, the entropy calculations have shown that we should take seriously
the interpretation of $S_{\ssc BH}$ as a statistical mechanical entropy.
Hence, with a pure state, since the entropy vanishes,
one must have that $A_{\ssc H}=0$, \ie there is no horizon and no black hole.
In the D-brane calculations, it is only by considering a mixed state including
all of the degenerate configurations that the black hole results
are matched.

The requirement of a mixed state 
becomes even more explicit in the scattering calculations. In order
to produce the correct black hole correspondence for,
\eg the absorption cross-section \cross,
one calculates with a decoherent ensemble of all
degenerate D-brane configurations. In particular, any generic tampering with
the low energy behavior $\langle n_{\omega/2}\rangle\simeq1/\beta\omega$
would destroy the correspondence to the universal black hole
cross-section. 

The precise 
connection between D-brane configurations at weak-coupling and
 gravitational field solutions at strong-coupling is
not well understood. Certain qualitative aspects are clear, \eg
these configurations develop 
strong gravitational fields over a range much larger than the string scale.
The original expectation was that, having prescribed a given set of 
asymptotic charges,
there is a unique black hole solution with those same charges and 
any of the corresponding D-brane configurations will
yield this same black hole.

While the uniqueness of the black hole solution may be true -- although
it remains to be rigorously proven -- a given set of charges simply do not
fix a unique field configuration. Actually, there are infinite families of
solutions, which, however, contain singular
null surfaces which are not concealed behind any event horizons
\garya\bright. Given that pure (or even generic mixed)
states of D-branes do not appear in strong-coupling to
correspond to a black
hole, it is thus clear that these non-black solutions provide a 
vast collection of alternatives.

Since the prospect of dealing with naked singularities must
seem an unsavory prospect to most relativists, several comments are in order.
First, when dealing with a system composed of a single type of D-brane, it is 
accepted
that the corresponding strong-coupling solutions have such
singular surfaces.
One expects in a certain sense that string theory is telling
us about the resolution of these singularities with D-branes.
Further, when I say there are ``singularities'' present,
I mean the solutions
of the leading-order low-energy string equations are singular.
The full string equations will modify the strong curvature regions, and
at the ``singularity'' the full stringy nature of these
configurations should become manifest. There it is likely
that even the notion of a spacetime
metric must be discarded.

I would add that further interesting evidence in support of
these arguments can be found in \garya. There, in an ensemble
of D-brane configurations, the contributions are weighted
in order to consider a nonuniform charge density. Hence
the degeneracy is reduced, but this is precisely
matched by a reduction of the horizon area, maintaining the
expected equivalence $S_{\ssc BH}=\ln\D$. However, one finds that 
even this benign tampering with the mixed state
produces a mild curvature singularity at the horizon\garyb. 

Given that pure states do not form black holes, it may seem
that only exceptional circumstances can lead to
black hole formation within a quantum framework.
This is, of course, not the case. Rather only the standard
mechanisms of decoherence are required\dec.
As a practical matter, if some segment of a quantum system is considered,
a pure initial state can appear 
to become a mixed state through random interactions with
its environment. For example, in the D-brane model, if
a given observer is unable or chooses
not to observe low energy radiation below some energy scale $\omega$,
one can estimate from \cross\ that a mixed state forms 
on a decoherence time-scale $T^{-1}\buildchar{>}{}{\sim}
\omega^7(A_{\ssc H}\beta)^2$ (in four-dimensional spacetime).
For a fixed $\omega$, the decoherence time-scale becomes especially short
at strong-coupling. Thus, as a practical matter,
pure D-brane states rapidly evolve to mixed states, which would
then display the characteristics of a black hole.
There must be strong correlations, however, between the ``black hole
state'' and its environment  even at its inception.

To summarize, then, recent progress in understanding black hole
microphysics carries a simple message: pure quantum states do not
form black holes. Rather, such states are likely to correspond
to field configurations with ``singularities'', strong
curvature regions, which are not clothed by horizons. 
Thus the information paradox results from an ill-posed question.
In completely understanding black hole evaporation, one is still
likely to face information loss questions. However, the discussion
must now be formulated in a framework where black holes are inherently
associated with mixed states. In this approach, the resolution of
these questions seems likely not to require new physics
incorporating information
loss as a matter of principle. Rather, it seems likely that the
information loss arises as a common  matter of practice,
with which we are familiar in our everyday lives.

\bigskip
I thank James Anglin and Ted Jacobson for useful comments, and
Ramzi Khuri for suggestions on improving the manuscript.
I would also like to thank Peter Haagensen for his efforts
in submitting my essay on time.
This research was supported in part
by NSERC of Canada and Fonds FCAR du Qu\'ebec.

\listrefs
\end